\documentclass[preprint, journal]{vgtc}       

\ifpdf
  \pdfoutput=1\relax                   
  \usepackage{graphicx}                
  \DeclareGraphicsExtensions{.pdf,.png,.jpg,.jpeg} 
\else
  \usepackage{graphicx}                
  \DeclareGraphicsExtensions{.eps}     
\fi%

\graphicspath{{figures/}{pictures/}{images/}{./}} 

\usepackage{hyperref}

\usepackage{microtype}                 
\PassOptionsToPackage{warn}{textcomp}  
\usepackage{textcomp}                  
\usepackage{mathptmx}                  
\usepackage{times}                     
\usepackage{cite}                      
\usepackage{tabu}                      
\usepackage{booktabs}                  

\usepackage{enumitem}

\usepackage[dvipsnames]{xcolor}
\usepackage[normalem]{ulem}
\definecolor{mred}{rgb}{.80,.12,.30}
\definecolor{grey}{rgb}{0.5,0.5,0.5}
\definecolor{lgrey}{rgb}{0.7,0.7,0.7}
\definecolor{purple}{rgb}{.75,0,.85}
\definecolor{pistachio}{rgb}{0.58, 0.77, 0.45}
\definecolor{myorange}{rgb}{0.94, 0.36, 0.13}

\newif\ifnotes
\notestrue

\let\origcite\cite
\renewcommand{\cite}[1]{\ifnotes\mbox{\origcite{#1}}\else \origcite{#1}\fi}

\preprinttext{\emph{2022 IEEE Workshop on Visualization Guidelines in Research, Design, and Education (VisGuides)}}

\onlineid{1010}

\vgtccategory{Research}

\title{Analysis Without Data: Teaching Students to\\ Tackle the VAST Challenge}

\author{Edward W He, Daniel Tolessa, Ashley Suh, Remco Chang}
\authorfooter{
\item
Edward W He, Ashley Suh, and Remco Chang are with Tufts University. E-mails: \{edward.he, ashley.suh, remco.chang\}@tufts.edu.
\item
 Daniel Tolessa is with Virginia Tech. E-mail: danielmt@vt.edu.
 }

\shortauthortitle{He \MakeLowercase{\textit{et al.}}: Analysis Without Data: Teaching Students to Tackle the VAST Challenge}

\abstract{
The VAST Challenges have been shown to be an effective tool in visual analytics education, encouraging student learning while enforcing good visualization design and development practices.
However, research has observed that students often struggle at identifying a good ``starting point'' when tackling the VAST Challenge.
Consequently, students who could not identify a good starting point failed at finding the correct solution to the challenge.
In this paper, we propose a preliminary guideline for helping students approach the VAST Challenge and identify initial analysis directions.
We recruited two students to analyze the VAST 2017 Challenge using a hypothesis-driven approach, where they were required to pre-register their hypotheses prior to inspecting and analyzing the full dataset.
From their experience, we developed a prescriptive guideline for other students to tackle VAST Challenges.
In a preliminary study, we found that the students were able to use the guideline to generate well-formed hypotheses that could lead them towards solving the challenge.
Additionally, the students reported that with the guideline, they felt like they had concrete steps that they could follow, thereby alleviating the burden of identifying a good starting point in their analysis process.

} 

\teaser{
  \centering
  \includegraphics[width=.9\linewidth, trim={10pt 160pt 20pt 0}, clip]{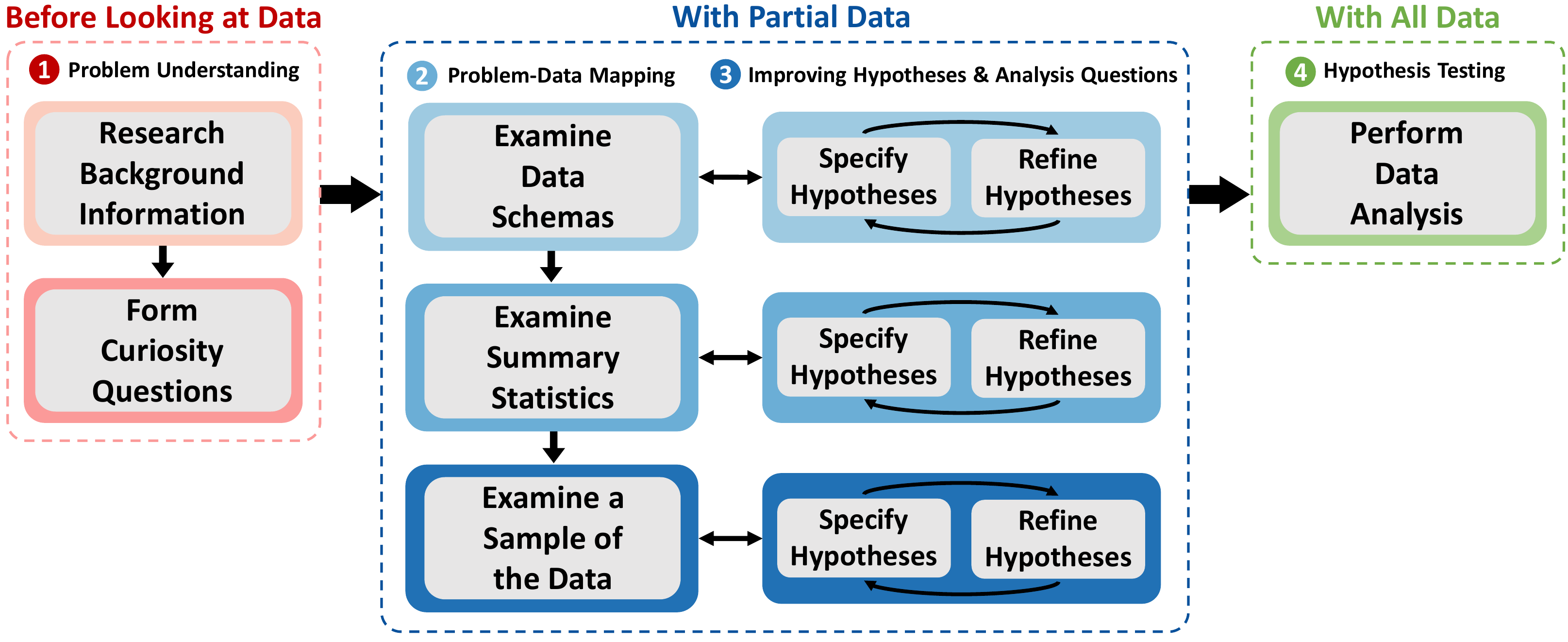}
  \caption{Our four-stage proposed guideline, HDAV (\textbf{H}ypothesis-\textbf{D}riven \textbf{A}pproach for \textbf{V}AST), for students tackling VAST Challenges. HDAV follows the practice of pre-registering hypotheses before visual analysis and experimental design\cite{hullman2021designing, cockburn2018HARK}, encouraging students to critically consider the \textit{analytic questions} prior to analyzing the full dataset. Only after a student has fully articulated their hypotheses should they move to performing data analysis. The final arrow denotes that once the data analysis process begins, students should not return to forming hypotheses. Doing so could result in HARKing (Hypothesizing After Results are Known)\cite{kerr1998harking} and false discoveries\cite{zgraggen2018investigating}.}
  \label{fig:model}
}

\vgtcinsertpkg

\begin{document}


\firstsection{Introduction}

\maketitle

The VAST Challenges\footnote{\small \url{https://vast-challenge.github.io/2022/about.html}} emulate real-world visual analytic tasks and problems, making them an effective tool in visual analytics education.
In classroom settings, researchers and educators have found that the VAST Challenge enables situated learning, which emphasizes learning around realistic case studies and promotes exploration by the students, resulting in active learning of complex concepts\cite{whiting2009vast, rohrdantz2014augmenting}.
However, the same research has also found that the students' performance can vary because of a lack of clear starting points.
This results in the students exploring the data seemingly randomly and stumbling upon the right (and sometimes wrong) answers:
\begin{itemize}[topsep=1pt, partopsep=1pt,itemsep=2pt,parsep=2pt]
    \item[] {\textit{\ldots There is no clear starting point in the [Blue Iguanodon] scenario, so students must start by searching for anything ‘interesting.’ In some cases, students seemed to serendipitously stumble upon the key component of the answer, which then easily led to much of the rest of the answer. Thus, students either got most of the answer or none at all. (Whiting et al.\cite{whiting2009vast})}} 
\end{itemize}

Searching unmethodically for anything ``interesting'' can be costly and threaten the validity of a student's analysis.
While it is possible to discover interesting patterns in the data can represent the ground-truth, they can also be spurious and meaningless.
In a study with university students, Zgraggen et al.\cite{zgraggen2018investigating} found that over 60\% of the students' findings in open-ended data exploration with a visual analysis tool consist of false discoveries (Type I error where the found insights are wrong) or false omissions (Type II error where the correct answers are omitted).
Without guidance to identify better ``starting points,'' students may fall into these common pitfalls when solving VAST Challenges. 


In this paper, we present a preliminary guideline for helping students in a visual analytics classroom to overcome the initial steps of tackling a VAST Challenge.
Motivated by the above work\cite{zgraggen2018investigating, whiting2009vast}, our key insight in forming this guideline is to discourage students from immediately trying to explore the data.
Instead, we want to encourage students to spend time on the {\it analytics} aspect of the challenge -- focusing on reasoning through the problem and tasks before digging into the data. 

This approach of methodologically performing data analysis
is inspired by recent work in pre-registering hypotheses before visual analysis\cite{kim2019bayes, suh2022grammar} and experimental designs\cite{cockburn2018HARK}. 
In experimental studies, pre-registration can prevent p-hacking\cite{dragicevic2016fair}, HARKing (Hypothesizing After Results are Known)\cite{kerr1998harking}, and postdiction (as opposed to prediction)\cite{nosek2018preregistration}.
In exploratory data analysis, researchers have found that asking analysts to describe their hypotheses prior to performing data analysis encourages the analysts to investigate the data in more detail\cite{choi2019concept}, which can in turn result in deliberate analysis outcomes\cite{koonchanok2021data}. 

Based on these collective findings, we propose a Hypothesis-Driven Approach for VAST Challenges (HDAV). HDAV is designed to guide students through the initial understanding and reasoning processes of a VAST challenge, in order to thoughtfully form and refine hypotheses before conducting analysis. As evidenced by Whiting et al.\cite{whiting2009vast} and Zraggen et al.\cite{zgraggen2018investigating}, such an approach can prevent students from serendipitously finding the `correct' answer, jumping to spurious conclusions, or pigeonholing themselves into their first `interesting' finding. HDAV consists of four stages that reflect a \textbf{student's incremental access to the data for data analysis}.
These four stages include: (1) Problem Understanding, (2) Problem-Data Mapping, (3) Improving Hypotheses and Analysis Questions, and finally (4) Data Analysis.

We follow an iterative approach in developing the guideline.
In the first phase, two undergraduate students (who are co-authors of this paper) investigated the 2017 VAST Challenge Mini-Challenge 1\cite{whiting2017vast} following the pre-registration practice of articulating their analysis questions and hypotheses while minimizing their exposure to the raw data.
Similar to the process of {\it coding} in qualitative studies\cite{macqueen1998codebook}, the two students examined the VAST Challenge independently and met over multiple sessions to achieve internal agreement between their listed analysis questions, as well as their overall analysis approach.
In the second phase, the research team (including all co-authors of this paper) reflected on the students' process of formulating and refining analysis questions, resulting in our proposed guideline.
Finally, the same two undergraduate students applied the guideline to a new VAST Challenge (in this case, the 2016 VAST Challenge Mini-Challenge 2\cite{crouser2016vast}). 
Role-playing as potential students in a visual analytics course, the two students documented their experience and reported the benefits and the shortcomings of using the guideline.

We note that the guideline reported in this paper is preliminary. 
Future controlled experiments with more students in a visual analytics course is necessary to validate the efficacy of our guideline.
Therefore, we present our guideline as a series of concrete steps that can help students and educators reason about the {\it analytical} questions for a VAST Challenge independently from the data analysis (shown in Figure~\ref{fig:model}).
Our findings from using the guideline in the 2016 VAST Challenge suggests that students can benefit from a more methodological approach to starting analysis than ``searching for anything interesting''\cite{whiting2009vast}, and motivates future work for visual analytics education. 

\section{Related Work}
Our proposed guideline is based on past literature on the use of the VAST Challenge and the authors' experience with a university-level visual analytics course. 
The guideline differs from, but is consistent with, the existing models in visual analytics.
In this section we describe past work in these two areas and how our work relates to them.

\subsection{VAST Challenge in Education}
The IEEE Visual Analytics Science and Technology (VAST) Challenge was first introduced in 2006 as a contest to advance the science of visual analytics.
The VAST Challenges have evolved to serve as a benchmark for the visual analytics community to experiment with new interactive systems, algorithms, and evaluation techniques.
The Challenges are typically modeled after real-world scenarios and each challenge comes with a ``ground-truth'' solution\cite{cook2014vast}. 
Each challenge typically consists of three independent mini-challenges (MCs) that reflect different problem domains (e.g., temporal analysis, image analysis, text analysis, geospatial analysis, etc.).
Solving all three mini-challenges will reveal the higher-level ``story'' of the overall challenge\cite{cook2014vast, scholtz2012reflection}.

As a platform for evaluating visual analytics environments\cite{scholtz2014evaluation}, the VAST Challenge holds many benefits when used in an education context. 
Whiting et al.\cite{whiting2009vast} reported an early experiment at Virginia Tech where the VAST Challenge was used in two Computer Science courses (for graduate and undergraduate students respectively).
Using the 2007 VAST Challenge, the authors found it encouraged active learning and fostered critical thinking, as the concept of a ``ground-truth'' to be discovered by the students was an intriguing intellectual puzzle. Additionally, the realistic nature of the challenge motivated the students to apply what they had previously learned in class\cite{whiting2009vast}. 
These findings were confirmed by another study in 2013 at the University of Konstanz where the researchers used the Bloom's taxonomy of Educational Objectives and demonstrated how the VAST Challenge enabled the integration of different learning objectives\cite{rohrdantz2014augmenting}.

However, while both studies reported positive outcomes from using the VAST Challenge for education, they also noted potential difficulties and pitfalls. 
From the perspective of the instructor, the use of the VAST Challenge increased supervision effort and raised questions about how to design a consistent grading rubric. 
From the perspective of the students, both studies reported the importance and the high cost of the students taking a ``wrong direction,'' especially during the initial phase of the analysis. 
The goal of our paper is to provide a preliminary guideline that can help students avoid such pitfalls in the initial phase and increase their chances of successfully solving VAST Challenges.

\subsection{Visual Analytics Models and Guidelines}
There have been a number of models (or workflows) proposed for the visual analytics community.
These models have defined the process of visual analytics and identified opportunities to improve current practices.
Some of the more notable models include the visual analytics ``diamond'' model by Keim et al.\cite{keim2008visual}, Card and Pirolli's Sensemaking model\cite{pirolli2005sensemaking}, van Wijk's value of visualization model\cite{van2006views}, Sacha et al.'s knowledge-generation model\cite{sacha2014knowledge}, and the more recent models of visual analytics as (machine learning) model building and exploration\cite{andrienko2018viewing, cashman2019user}.

However, while these models have been invaluable to the visual analytics community, generally they have been designed for researchers and are ``descriptive'' in nature.
For students who are not well-versed in visual analytics literature, these descriptive models are insufficient in helping them identify a good ``starting point'' when addressing a complex visual analytics problem like the VAST Challenge. 

To develop a ``prescriptive'' guideline that is more suitable for students, we draw inspiration from the existing visual analytics models and the Design Study Methodology paper by Sedlmair et al.\cite{sedlmair2012design}.
The goal of the guideline is to have a linear process that a student can follow where each step of the process is associated with concrete activities and examples. 
Figure~\ref{fig:model} illustrates the workflow of our proposed guideline.
In the sections below, we describe the process for generating the guideline and the explanations for each of the steps.


\section{Methodology for Tackling the VAST Challenge}
\label{sec:methodology}
Our goal is to develop a guideline that students in a university-level visual analytics course can follow when working through a VAST Challenge. 
To do so, we recruited two student participants to complete a VAST Challenge, with specific constraints and criteria, and rigorously document their process.
The resultant guideline outlines a step-by-step workflow that includes: 
(1) concrete instructions for students to follow,
(2) specific examples from the 2017 VAST Challenge to provide clarity, and (3) additional suggestions for steps that students may have trouble completing. 
We describe the process for developing the guideline in terms of the material, the participants, the task, and the procedures.

\subsection{Material}
\label{sec:method:material}
The problem used for this task was the 2017 VAST Challenge\footnote{\small \url{http://vacommunity.org/VAST+Challenge+2017}} Mini-Challenge 1 (MC1). 
This challenge is based on a fictitious nature preserve where the local Red Crested Blue Pipet (bird) resides, and a (fictitious) problem in the preserve that must be solved: 
Recently, it has been discovered that the nesting pairs of the Blue Pipets have been decreasing. 
The local Ornithology Preservation Society is concerned about the decrease in these nesting pairs and would like to identify possible reasons for this. 
The goal of MC1 is to identify if there is a link between the traffic going through the park and the decrease in nesting pairs. 
The questions provided by the challenge are: 

\begin{itemize}[topsep=1pt, partopsep=1pt,itemsep=1pt,parsep=1pt]
    \item[\textbf{Q1}] Describe daily patterns of vehicles traveling in the park.
    \item[\textbf{Q2}] Describe patterns of vehicles traveling through the park over a longer period of time (multiple days).
    \item[\textbf{Q3}] Identify unusual patterns of activity.
\end{itemize}

MC1 data includes: (1) a map of the nature reserve (as an image), and (2) a tabular dataset consisting of sensor-log information for vehicles visiting and driving through the preserve. 
Each time a car passes through one of the sensor locations, the time-stamp, sensor-ID, car-ID, and car-type are recorded as a row in the data.
In addition to the data, MC1 includes a data description and a list of questions that a contestant is required to complete (referred to as the ``answer sheet'').

\subsection{Participants}
Two undergraduate students majoring in Computer Science (who are co-authors of this paper) served as the ``participants'' in tackling the 2017 VAST Challenge. The students were recruited as part of a summer undergraduate research appointment.
The two students had not previously taken a visual analytics course and had little experience with visual analytics systems (e.g. Tableau). As such, we consider these participants ``typical students'' with the average criteria of those who might be taking a visual analytics course at the university level. 

\subsection{Tasks}

\begin{table}[t]\centering
\small
\begin{tabular}{p{0.05\linewidth} p{0.8\linewidth}}
        \toprule
        \textbf{Level} & \textbf{Criterion} \\ 
        \midrule
        0 & No explanation provided. A nonsense statement, a question, an observation, or a single inference about a single event, person, or object. \\
        1 & An inappropriate explanation. For example, ``\ldots because it’s magic'' or ``\ldots because the man pushed a button.'' \\
        2 & Partial appropriate explanation. For example, incomplete reference to variables, a negative explanation, or analogy. \\
        3 & Appropriate explanation relating at least two variables in general or nonspecific terms. \\
        4 & Precise explanation with qualification of the variables. A specific relationship between the variables is provided. \textbf{This is a hypothesis}. \\
        \bottomrule
    \end{tabular}
    \caption{\small Sunal and Haas' \textit{Hypothesis Quality Scale}\cite{sunal2002social} for writing and evaluating students' hypotheses. Level 4 and beyond is considered a valid hypothesis.%
    \vspace{-10pt}
    }
\label{tab:hyp_quality_scale}
\end{table}

The two participants were tasked with approaching the VAST Challenge with the following constraints:

\begin{itemize}[leftmargin=0pt,topsep=1pt, partopsep=1pt,itemsep=1pt,parsep=1pt]

    \item[] \textbf{Generate Analytic Questions and Hypotheses}: Following the concept of pre-registration\cite{cockburn2018HARK}, the two participants were instructed to generate analysis questions or hypotheses related to the challenge before performing data analysis. Questions and hypotheses were specified to be related to Q1-Q3 from the challenge. For example, a question could be on what could qualify as an ``unusual pattern of activity'' (MC1 Q3). 
    
    \item[] \textbf{Limit the Use of Raw Data}: To mitigate potentially spurious findings\cite{zgraggen2018investigating}, the participants were asked to generate as many analytic questions as possible before looking at the available data. After exhausting all relevant questions, they were then permitted to look at parts of the raw data. For example, the participants could look at the schema of the data or a few rows of the data. Each time the participant looked at some aspect of the data, they were instructed to generate additional questions and hypotheses, but \textit{not} attempt to validate them. This process was repeated until participants reported they had done all they could with hypothesis generation without performing data exploration or analysis.
    \item[] \textbf{Turning Open-ended Questions into Formal Hypotheses}: To ensure that the participants were forming meaningful analytic questions and hypotheses that could eventually be tested to solve the challenge, the participants were asked to refine any open-ended questions into formalized hypotheses using either the hypothesis grammar by Suh et al.\cite{suh2022grammar} or the Hypothesis Quality Scale by Sunal and Haas\cite{sunal2002social} (see Table~\ref{tab:hyp_quality_scale}). An example of this process is discussed in Section~\ref{sec:refinement}.
    \item[] \textbf{Careful Documentation}: Each participant was asked to keep a detailed research journal of: (1) the questions or hypotheses that they generated, (2) parts of the data they examined, (3) tools that they used to examine the data (e.g., Excel, Tableau), (4) how the data was examined (e.g., screenshots of tables or visualizations), and (5) additional notes and observations about their process. 
    \item[] \textbf{Iterative and Collaborative Refinement}: The two participants were asked to complete all components of the task independently. After both had completed a round of forming analytic questions for the challenge, they were asked to meet with the other person and compare their research journals. After each meeting, the participants were asked to return to the task and refine their questions and hypotheses based on the previous discussion. This iterative process continued until the two participants agreed that their two lists of questions and hypotheses were sufficiently refined, exhaustive, and consistent with each other's. 

\end{itemize}

\subsection{Procedure}
The two participants (henceforth referred to as PID1 and PID2) completed the task over the span of 2 weeks.
The participants initially took different approaches.
After reading the MC1 challenge (and without looking at the data), PID1 began the task by performing background research into general bird nesting behavior (e.g., by reading online articles) while PID2 immediately started posing questions and hypotheses.

It took the participants three iterations before they agreed that their two lists of questions and hypotheses were consistent with each other's.
For example, during the discussion after their first iteration, the participants noted that PID1's questions were strictly focused on unusual patterns of activity (MC1 Q3), and did not generalize to the normal behavior of vehicle activity. On the other hand, PID2's questions addressed each component of MC1, but were much broader than PID1's questions. Therefore, in the next iteration of the task, PID1 spent time generating questions for the remainder of MC1 while PID2 spent time refining his questions to better categorize unusual activity. 

The two varied methods in which the participants posed questions is consistent with prior literature where researchers documented the different ways that students approach the VAST Challenge\cite{whiting2009vast}. 
Over the course of two additional iterations, the participants were able to come to an agreement that their lists of questions and hypotheses were the same (or sufficiently similar) in both coverage and specificity.

The participants consolidated their research journals to construct the initial draft of the guideline.
This guideline underwent further editing and refinement by the research team (the co-authors of this paper).
In the section below we describe our proposed guideline for students to approach a VAST Challenge, using the iterative process conducted and fine-tuned by the task participants.

\section{The Guideline}
The guideline aims to provide a structured step-by-step process to guide students towards solving a VAST Challenge. 
With the focus on encouraging the students to critically analyze the problem before performing data exploration and analysis, this guideline consists of 3 parts: (1) problem understanding, (2) problem-data mapping, and (3) hypothesis formation.
Figure~\ref{fig:model} illustrates the workflow of the guideline.

\subsection{Before Looking at the Data: Problem Understanding}
\label{sec:problemUnderstanding}
The first stage of the guideline is to carefully review all materials of the VAST Challenge and brainstorm about the possible hypotheses that can be formed, without looking at or analyzing the data.
VAST Challenges come with a wealth of contextual information in addition to the raw data:
A typical challenge includes a background story about the scenario, an ``answer sheet'' that contains the questions that a contestant needs to complete, and (sometimes) data dictionaries that explain the dataset.
Even before looking at the data, a student can begin to critically reason about the challenge and form questions and hypotheses that will guide the eventual solution.

\subsubsection{Background Research}
In order to have a good understanding of the problem, students should perform some background research to familiarize themselves with the story in the challenge.
For example, understanding the roles of a park ranger in a US nature reserve will help a student better assess the behavioral patterns of the ranger vehicles in the 2017 Challenge.

Most critically, the students need to carefully review the questions in the ``answer sheet.''
Often the answer sheet provides hints to the possible solution. 
For example, Q1 of the 2017 Challenge MC1 states:
\begin{itemize}[topsep=1pt, partopsep=1pt,itemsep=2pt,parsep=2pt]
    \item[] {\it Characterize the patterns by describing the kinds of vehicles participating, their spatial activities (where do they go?), their temporal activities (when does the pattern happen?), and provide a hypothesis of what the pattern represents (for example, if I drove to a coffee house every morning, but did not stay for long, you might hypothesize I’m getting coffee ``to-go'')}
\end{itemize}
This question indicates that finding the solution requires a geospatial-temporal analysis grouped by vehicle types.
Without looking at the data, the student can reasonably assume what the data will contain (e.g., timestamps for different vehicles across regions of the park) and the types of visualizations (e.g., time series charts or choropleths) that would be helpful for solving the challenge.

\subsubsection{Curiosity Questions}
\label{sec:guide:curiosity}
After reviewing the background material, the students should write down any initial questions 
inspired by the challenge. We refer to these as ``curiosity questions.'' Curiosity questions are questions that students can pose about a scenario without any restriction or concern on what the answer \textit{ought} to be. 
This step is beneficial as it aids with brainstorming and forming ideas without worrying about the quality or the validity of those questions. 
In a collaborative classroom setting, it would also encourage team members to adjust to each other's prior beliefs about the scenario. 
Example curiosity questions generated by the two participants include:
\begin{itemize}[topsep=1pt, partopsep=1pt,itemsep=1pt,parsep=1pt]
    \item What time do cars usually enter the park?
    \item How many less nesting pairs are there?
    \item Is there an increase in noise disturbance?
    \item Where in the park are birds most active?
    
\end{itemize}

\subsection{Peeking at the Data: Problem-Data Mapping}
\label{sec:peeking}
Once the students have understood the scenario of the challenge and posed their curiosity questions, the next step is to map the questions to the data.
Similar to the concept of pre-registration, the students will not have full access to the raw data when generating analysis questions and initial hypotheses. 
Instead, the students will have step-by-step incremental access to partial data (we refer to this as ``peeking'' at the data), in which they examine: (Step 1) the data schema, (Step 2) summary statistics, and (Step 3) a sample of the data.


\subsubsection{Step 1: Examining the Data Schemas}
During the initial step, students are allowed to look at the data schemas. 
Data schemas provide essential information about the dataset, such as the number of columns, their names, and their data types.
Examining the schemas will help the students recognize which of their curiosity questions can actually be answered by the data that is provided. 

For example, in Section~\ref{sec:methodology}'s exercise, the participants were tasked with discovering the cause of the decrease in Pipets' nesting while completing the 2017 VAST MC1. 
Unsurprisingly, most of the curiosity questions posed by the participants related to causal reasoning about the nesting decrease (examples provided in Section~\ref{sec:guide:curiosity}).
However, upon examining the data schema, the participants recognized that no nesting information is provided in the data -- the data only contains sensor information about traffic in the nature reserve (see Section~\ref{sec:method:material}).
At this point, the participants refined their initial questions and switched from asking causal questions to asking questions about detecting anomalies, or abnormal patterns, in the data.


\subsubsection{Step 2: Examining the Summary Statistics}
Summary statistics are used to provide a holistic view of the data as simply as possible. 
In Section~\ref{sec:methodology}'s exercise, we constrained summary statistics to include central tendency (mean, mode, median), spread (standard deviation, range, percentile), shape (distribution), in addition to the characteristics and structure of the data: cardinality, number of rows, and completeness (if there are missing values). 

In the second step of peeking at the data, the students should review the next draft of their questions with the summary statistics in mind.
The students might realize that the size of the data is very large and scalability could become a concern, or that there are missing values in the data that need to be addressed.
Most importantly, students should recognize whether their questions might not be answerable without performing some kind of data transformation.
Data transformation could require transforming the values within the same column (e.g., normalizing the data to fit the range of [0, 1]), performing data joining (e.g., merging two tables in a database into a single table), or deriving new columns entirely.

For example, to complete the 2017 VAST MC1, a question pertaining to whether cars are speeding in the preserve will require calculating the distances between sensors -- which can only be done by examining the map -- and the elapsed time between two consecutive sensor readings of the same car.
At the end of this step, students should have an updated list of questions and potential hypotheses, as well as their strategies for data transformation that are necessary to answer their questions.

\subsubsection{Step 3: Examining a Sample of the Data}
During the third step, students should have access to a sample of the data.
The purpose of this step is to do a ``trial run'' of the full analysis experience so that the students can: (1) generate initial visualizations and analysis scripts based on the sample data, and (2) assess whether the tools that they have are sufficient for their analysis needs.

As an example, in the 2017 VAST MC1, the students should recognize that the map (as an image) will be difficult to work with, but analysis of which will be important.
Road information (e.g., connectivity and distances between sensor locations) can only be extracted by examining the map.
However, as there is no standard mechanism to extract such information, the students should begin to think about whether they want to manually compute the needed data or write image-processing tools to do automatic extraction.

At the end of this step, the student should have a better sense of how their questions and hypotheses can be addressed with the data. They should also be able to document the tools that are necessary to complete their analysis and estimate the difficulty of completing remaining tasks.

\subsection{Improving Hypotheses and Analysis Questions}
\label{sec:refinement}
Most undergraduate students in a visual analytics course do not have formal training in business or intelligence analysis, which is often a strong component in solving VAST Challenges. 
Without such training, students can have difficulty coming up with ``good'' analysis questions and thus testable hypotheses. 

For example, in a previous visual analytics university course where students were given the 2017 VAST MC1, a student submitted the following hypothesis to Q3: {\it ``certain visitors are irregularly active in certain regions of the park, causing the decreased nesting.''}
Although this hypothesis matches the problem description (a geospatial-temporal analysis of vehicles), this hypothesis lacks specificity for further analysis (i.e., who are the {\it ``certain visitors,''} what are the {\it ``certain regions,''} and how to define {\it ``irregular active?''}).

The purpose of this stage is to help the students generate, assess, and refine their analysis questions and hypotheses. 
As shown in Figure~\ref{fig:model}, this exercise happens at the end of each step of the Problem-Data Mapping stage.
The goal is to ask the students to reflect on their analysis questions and refine them into formalized hypotheses before moving to the next step in the process.

\subsubsection{Hypothesis Specification}
One way to assess the quality of an analysis question is to formalize the question as a scientific hypothesis.
A valid scientific hypothesis states a qualified or quantified relation between two variables and is falsifiable (i.e., it has a binary true or false outcome)\cite{popper2005logic}.
Rich literature has been contributed on assessing and refining the quality of a hypothesis from the science education community\cite{kroeze2019automated, mulder2010finding, van1991supporting}.
For the purpose of our project, we use the Hypothesis Quality Scale proposed by Sunal and Haas\cite{sunal2002social} (shown in Table~\ref{tab:hyp_quality_scale}).

To help students with the hypothesis specification process, consider the following example where the student might start with the following analysis question:
\begin{itemize}[topsep=1pt, partopsep=1pt,itemsep=1pt,parsep=1pt]
    \item {\it When do cars generally enter the park?}
\end{itemize}
This represents a Level-0 hypothesis in the Hypothesis Quality Scale in that it is a question without explanation. 
An improvement over this question might be:
\begin{itemize}[topsep=1pt, partopsep=1pt,itemsep=1pt,parsep=1pt]
    \item {\it Do cars enter the park mostly in the mornings?}
\end{itemize}
This question might represent a Level-2 hypothesis in that it has a true or false answer, but it is incomplete in its references.
Another iteration of improvement results in:
\begin{itemize}[topsep=1pt, partopsep=1pt,itemsep=1pt,parsep=1pt]
    \item {\it There are more cars entering the park between 5-8am than any other three hour period.}
\end{itemize}
This hypothesis is testable and specific, representing a Level-4 hypothesis in the Quality Scale.

\subsubsection{Hypotheses Refinement}
\label{sec:hyprefine}
Representing an analysis question formally as a hypothesis has the additional advantage that it can be further refined.
For example, starting with the previous hypothesis, a student can now alter parts of the hypothesis to afford more flexibility in the analysis:
\begin{itemize}[topsep=1pt, partopsep=1pt,itemsep=1pt,parsep=1pt]
    \item {\it There are more cars entering the park between 5-\textcolor{blue}{9}am than any other \textcolor{blue}{four} hour period.}
\end{itemize}

In addition to changing the value ranges, the students can also add additional specifications:
\begin{itemize}[topsep=1pt, partopsep=1pt,itemsep=1pt,parsep=1pt]
    \item {\it There are more \textcolor{blue}{2-axle} cars entering the park between 5-9am than any other \textcolor{blue}{car type} in a four hour period.}
\end{itemize}

Students are encouraged to reflect on the hypothesis refinement process with respect to the VAST Challenge questions.
In the case of the 2017 VAST MC1 Challenge -- which hints at a geospatial-temporal analysis grouped by car-types -- the students should investigate refining their hypotheses along those three dimensions.

\subsection{Data Analysis}
The last step of the guideline is for the students to perform their planned data analyses in solving the VAST Challenge using the hypotheses that they have developed in the previous stages.
In some courses, the students use existing tools to solve the challenge (e.g., the courses at Virginia Tech\cite{whiting2009vast}) while other courses require the students to develop novel visual analytics systems that are tailored for the challenge (e.g., the visual analytics course at University of Konstanz\cite{rohrdantz2014augmenting}). 

Regardless of the structure of the course, it is important to note that once the students start the data analysis process, it is not advisable for them to return to the early stages of the guideline. 
Posing research questions and hypotheses after data analysis runs the risk of HARKing and Type I errors (false discoveries) in the analyses\cite{zgraggen2018investigating}. 

Whether a course instructor should strictly disallow students from back-tracking depends on the course's learning objectives. 
One could argue that back-tracking should be allowed, as students should be free to make mistakes in a classroom setting as long as they recognize these mistakes and can correct them.
Regardless of the instructor's decision on this matter, our guideline makes clear of the danger of back-tracking and strongly discourages the students from starting the data analysis process before being certain of their hypotheses.

We note that our guideline does not specify how data analysis should be conducted, instead, specifics should be justified by the student team in a write-up or predetermined by the course instructor. We discuss future work in which the guideline can be extended to include performing data analysis for students in Section~\ref{sec:future}.

\section{Applying the Guideline}
We perform a preliminary assessment of our guideline by applying the proposed workflow to a different VAST Challenge.

\smallskip
\noindent \textbf{Participants:} 
The same two students who analyzed the 2017 VAST Challenge, described in Section~\ref{sec:methodology}, participated in this exercise.

\smallskip
\noindent \textbf{Material:}
We use the 2016 VAST Challenge Mini-Challenge 2 (MC2).
This challenge presents a fictitious situation set on the island of Kronos where the company GAStech has just moved into a new building. 
The new building is equipped with sensors that detect a variety of building characteristics, such as temperature and chemical concentrations. 
GAStech has also implemented a new security procedure that requires staff to carry {\it prox cards}, which are read by mobile and stationary sensors located throughout the building and logged in a database. 
The goal of this challenge is to identify both typical and atypical patterns or areas of concern, given two weeks of building and prox card data. 

We note that the 2016 VAST MC2 is more complex than the 2017 VAST MC1.
Instead of only one tabular dataset with one map, the 2016 MC2 contains 8 tables, 10 JSON documents, and 9 map figures.

\smallskip
\noindent \textbf{Task:} 
The two participants 
were tasked with analyzing the 2016 Challenge MC2 independently (no communication occurred between the two until the end of the task).
Similar to the previous exercise, the participants were required to document detailed notes on their analysis and hypothesis refinement process -- but in this case, their approach succinctly followed the proposed guideline.
The participants were only required to reach Step 3 (examining data samples) of Problem-Data Mapping (Section~\ref{sec:refinement}), and did not perform any data analysis in the completion of their task.

\smallskip
\noindent \textbf{Outcomes:} 
The two participants spent approximately 8 days to complete this task (compared to 2 weeks in the previous exercise).
They were consistent in the time spent in each stage of the guideline.
Both spent 1-2 days generating their curiosity questions, 1-1.5 days between Step 1 (examining data schemas) and Step 2 (examining summary statistics) of Problem-Data Mapping, and 2-3 days in Step 3 (examining data samples).
In total, the participants spent 2-3 days specifying and refining their analysis questions into formalized hypotheses.

\section{Outcomes and Lessons Learned}
Using our guideline to solve the 2016 VAST MC2 provided a preliminary assessment of how this guideline could be helpful for future visual analytics students. 
On the positive side, the two participants reported that using the guideline resulted in a clearer and more streamlined experience where each step of the analysis process was concretely laid out.
However, the two participants also noted areas for improvement, which we document in this section.

\subsection{Quality and Quantity of the Generated Hypotheses}
We first examine the quantity and quality of the hypotheses generated by the two participants (PID1 and PID2 respectively).

PID1 generated 24 curiosity questions after conducting initial research into the challenge. After peeking at the data in Steps 1 and 2, PID1 added 18 new analytic questions. 
PID1 was then able to form 20 hypotheses after Step 3, and finally formalized 29 total hypotheses after refinement. However, PID1 noted that many of the 29 hypotheses have ``specifications'' that can be exchanged to pose different hypotheses (examples shown in blue in Section~\ref{sec:hyprefine}).
PID2 started with 8 curiosity questions after finishing initial research into the challenge. After peeking at the data in Steps 1-3, PID2 had 14 additional questions. In total, PID2 generated 15 hypotheses from 22 analytic questions.

Both participants generated hypotheses that could lead to discovering the ground truth (or valid answer) to the challenge.
For example, a ground truth provided in the 2016 VAST MC2 solution is that delivery workers enter/exit through the docks. A relevant hypothesis our participants came up with was ``There is a staff member who starts/ends the day not on floor 1, zone 1 (the main entrance).'' An evaluation of this hypothesis as true will help reveal delivery members that follow an ``atypical pattern,'' as described by the challenge.

Another ground truth (or valid answer) provided in the solution sheet is that hazium concentration spikes on F3Z1 on June 3 at 7AM, as well as on June 11 at 8PM. A relevant hypothesis formed by the participants was ``Between 12-1PM on 5/31 there is an entry for hazium concentration over 0.'' In this case, ``Between 12-1PM'' can be exchanged for any 1-hour period and ``5/31'' can be exchanged for any date. An evaluation of these hypotheses, using T as the correct time frame, will help the participants identify hazium concentration spikes.

Although our participants did not perform any data analysis to complete the 2016 VAST MC2, our preliminary findings suggest that, by following the guideline, future students will be able to identify good ``starting points'' for further investigation while following best practices in pre-registration and p-hacking prevention.

\subsection{Positive Lessons Learned}
The two participants reported three ways that the guideline helped with their analysis task.

\smallskip
\noindent \textbf{A Streamlined Process:} 
The participants found that the guideline provided a streamlined process for them to follow, resulting in time saved by the task being simpler to tackle. 
For the 2017 VAST Challenge, both participants had difficulty knowing where to begin analysis after reading the questions -- one started doing related research, while the other began posing questions. 
In the second task with the 2016 VAST Challenge, the procedures outlined in the ``Problem Understanding'' stage (Section~\ref{sec:problemUnderstanding}) was helpful in prescribing a place to start.

\smallskip
\noindent \textbf{Concrete Directions for Each Step:}
The participants noted that the guideline provided concrete directions for each step of the analysis process. 
Since ``peeking at the data'' (Section~\ref{sec:peeking}) was described to have three steps between examining the data schemas, the summary statistics, and a data sample, it was easier for the participants to know when and how much of the data they were allowed to access.
The clear distinction between the three steps also prevented the participants from looking at the full dataset before they had the chance to reflect on all possible questions, hypotheses, and analysis tasks.

\smallskip
\noindent \textbf{Guidance for Hypothesis Specification and Refinement:} 
The participants found the guidance for hypothesis specification and refinement (Section~\ref{sec:refinement}) to be useful in two ways.
First, following the process resulted in clearer hypotheses that were easy to share and communicate with others.
Second, with the use of the formal hypotheses, it was easier for the participants to consider alternatives via refinement.
For example, the participants were able to make the hypotheses more or less specific by considering different subsets of the data.

\subsection{Areas for Improvement}
While the two participants were positive about the guideline overall, they also noted two areas where it is incomplete or can be improved.

\smallskip
\noindent \textbf{How Many Hypotheses are Enough?} 
The 2016 VAST MC2 is more complex than the 2017 VAST MC1: 
2017 VAST MC1 contains only one tabular dataset, while the 2016 VAST MC2 has 8 tables that need to be joined prior to analysis.
As a result of the increased data complexity, the participants found that the guideline's suggestion of ``exhaustively generate as many hypotheses as possible'' was not attainable.
A practical method to help students decide when they can move on to the next step will be necessary to address the issue of data complexity.

\smallskip
\noindent \textbf{How Good are the Hypotheses?} 
Related, if the students cannot realistically generate all hypotheses, they need to be provided a way to rank the hypotheses and determine which ones to keep.
A strategy that the two participants used to combat this challenge was to group similar analysis questions and hypotheses into semantically meaningful clusters.
In future iterations of this guideline, we will look to incorporate this as a possible strategy for students to follow.

\section{Limitations and Future Work}
\label{sec:future}
We acknowledge that this guideline is only at a preliminary phase and is not ready for wide adoption without further considerations. 
As the guideline is generated by the same two students who tested the guideline, the lessons learned can be biased by their personal experience.
For example, although the two students were much faster in solving the second VAST Challenge using the guideline (8 days compared to 14 days), the speed-up could be due to their experience and not the guideline.
Similarly, we presume that there are other areas for improvement. 
The two points raised by the students both relate to the applicability of the guideline with regards to the changes in data complexity.
We imagine that if the guideline was used by a student who has never seen a VAST Challenge before, they would likely have additional suggestions or barriers faced.

In future work, we plan to refine the guideline by incorporating examples and lessons learned from the 2016 VAST MC2 exercise. Additionally, we will add prescriptive details on performing analysis given the students' completion of their hypotheses (i.e., following Section~\ref{sec:refinement}).
Further documentation for the guideline will be required to prepare it for dissemination to students in a visual analytics course.
Finally, we plan to conduct a controlled study in a visual analytics course that the co-author teaches at Tufts University in Spring 2023 to assess the effectiveness and limitations of the guideline in helping students tackle the VAST Challenge.



\section{Discussion}
\label{sec:discussion}
Although our proposed guideline is preliminary, our experience in developing the guideline has raised some interesting questions related to how VAST Challenges should be approached and whether the guideline can be generalized to visual analytics tasks beyond the use of the VAST Challenge in a classroom setting. 

\smallskip
\noindent \textbf{Data-Driven or Hypothesis-Driven?} 
While there has been no other guidelines in teaching students how to solve the VAST Challenge, past research seems to suggest that the students begin addressing a challenge by first performing data exploration (e.g.,\cite{whiting2009vast}).
Through the interactive exploration, the students form insights, come up with hypotheses, and use the visual analytics tool to refine and evaluate those hypotheses.

Our proposed hypothesis-driven guideline to solving the VAST Challenge is a departure from this approach.
We are encouraged by our preliminary finding that the students were able to generate hypotheses that were on the right track to discovering the ground truth without having access to the full dataset.
However, we note that this positive finding is not unique and has been studied with similar success in prior research\cite{choi2019concept, koonchanok2021data, kim2019bayesian, hullman2021designing, jun2019tea}.
%
Our work adds to this growing area of research and suggests the possibility of using a hypothesis-driven approach, as opposed to data-driven, for education purposes.

\smallskip
\noindent \textbf{Generalizability Beyond the VAST Challenge?} 
We consider the possibility of generalizing the proposed guideline to help analysts perform visual data analysis beyond the VAST Challenge.
VAST Challenges are modeled after real-world scenarios and have been touted to reflect intelligence analysis in practice\cite{scholtz2012reflection, cook2014vast}. 
However, we observe that there are also differences that could prevent the guideline from being applicable to general data analysis tasks.
First, VAST Challenges typically provide questions that hint at the final solution. 
In the 2017 VAST MC1, the questions indicate that the correct answer involves analyzing the data by first grouping the rows by car-types.
In real-world analysis tasks, it is not always the case that such instructions are available.

Second, real-world data are often much more complex than what is given in VAST Challenges.
When switching from the 2017 VAST MC1 to the 2016 VAST MC2, we observed a significant increase in the number of possible hypotheses that the students could have formed.
With even larger and more complex data, our proposed hypothesis-driven guideline would likely reach its limit, as an analyst could not systematically and exhaustively generate all possible hypotheses.
We speculate that there are ways to mitigate this problem (e.g., by clustering the hypotheses as the students did).
We leave the challenge of generalization for our guideline as future work.

\section{Conclusion}
In this paper, we present a preliminary guideline to help students in a university-level visual analytics course tackle a VAST Challenge.
Our work is based on the observation that without guidance, students currently struggle with identifying a good ``starting point'' when tasked with analyzing VAST Challenges.
Our guideline is developed based on our experience solving the VAST 2017 Mini-Challenge 1 following a hypothesis-driven approach. 
After completion, we applied the guideline to the VAST 2016 Mini-Challenge 2 to assess its benefits and limitations.
The preliminary results from this exercise suggest that the guideline provides the students with a streamlined approach and concrete steps to follow, reducing the students' time spent in the early phase of the task by identifying good analysis directions to pursue.

\acknowledgments{The authors wish to thank the NSF Diamond REU Program at Tufts University (2149871) and NSF grants 1855886, 1940175, 1452977, 1939945, 2118201 for supporting this work.}

\bibliographystyle{abbrv-doi}

\bibliography{template}
\end{document}